\documentclass[a4paper,11pt]{article}
\usepackage{graphicx}

\def\gto{\mathop {\hbox{${\lower3.8pt\hbox{$>$}}\atop{\raise0.2pt\hbox{$\sim$}}$}}}

\def\journaldata#1#2#3#4{{\it #1\/}\phantom{--}{\bf #2$\,$:} $\!$#3 (#4)}
\def\eprint#1{{\tt #1}}

\begin{document}

\title{Everpresent Lambda II: Structural Stability}

\author{
  {Maqbool Ahmed\footnote{Center for Advanced Mathematics and Physics,
      National University of Sciences and Technology, H-12 sector, Islamabad, Pakistan.
      email: mahmed@camp.nust.edu.pk}}
  and
  {Rafael D Sorkin\footnote{Perimeter Institute, 31 Caroline Street North, Waterloo ON, N2L 2Y5, Canada
      and
      Department of Physics, Syracuse University, Syracuse, NY 13244-1130, U.S.A.
      email: rsorkin@perimeterinstitute.ca}}}

\maketitle

\begin{abstract}
\noindent
Ideas from causal set theory
lead to a
fluctuating, time dependent cosmological-constant
of the right order of magnitude to match
currently quoted ``dark energy'' values.
Although
this effect
was predicted some time ago \cite{LamPred1,LamPred2},
it is only more recently that
a more detailed phenomenological model
of a fluctuating $\Lambda$
was introduced and simulated numerically \cite{everLam}.
In this paper
we continue the investigation by studying
the sensitivity of the model to
some of
the {\it ad hoc}
choices made in setting it up.
\end{abstract}

\section {Introduction}

As explained in reference [1],
a heuristic argument from causal set theory,
leads one to expect fluctuations
in the cosmological constant $\Lambda$
which scale inversely as the square-root of the spacetime volume.
(The reasoning rests on
the fundamental hypothesis of spatio-temporal discreteness
on one hand,
and on the quantal conjugacy\footnote%
{This conjugacy comes out very naturally in unimodular
 gravity \cite{unimod1,unimod2,unimod3}.}
or ``uncertainty relation''
between four-volume and $\Lambda$ on the other hand.)
Assuming that the fluctuations are centered on zero,
one obtains for the current epoch
the prediction $\Lambda\sim\,\pm 10^{-120}\rho_{Planck}$,
where $\rho_{Planck}$ is the Planck density.
It is thus natural to interpret
the observations
pointing to an accelerating Hubble expansion
as a confirmation of this prediction,
since the effective energy density
corresponding to the acceleration is of the same order of magnitude as
the predicted fluctuations.
(The sign of the fluctuations is purely random in this scenario, though
with a slight bias toward positive values, as described below.)


Not only does the current value of
$\Lambda$
receive a natural
explanation in this way,
but computer simulations of a simple
phenomenological
model \cite{everLam}
for the time-dependence of the fluctuations
have confirmed the suggestion that
a type of ``tracking'' behavior arises automatically,
in the sense that
the fluctuations
would have been
comparable to
the ambient matter density not only now, but in the past as well
(and also in the future, for as long as the expansion continues).
In this way the so called ``Why now?'' puzzle is also resolved.
(This puzzle has of course also called forth many other proposed
solutions, for example \cite{quint1,quint2,quint3}.  However most of them suffer from
the need for ``fine tuning''.)

The model
put forward in \cite{everLam}
derives $\Lambda$
(which can be interpreted quite generally as the action $S$ of free
spacetime per unit 4-volume)
from
a sum of random contributions to $S$
coming from the causal set elements
within the past light cone
of any given spacetime location.
It contains
a single phenomenological parameter
which reflects
both
the magnitude of the individual contributions to $S$,
and the conversion factor between
spacetime volume and number of causal set elements,
and which in the absence of ``fine tuning'' would have a value of order
unity.
Following \cite{everLam},
we will refer to this parameter,
which cannot at present be obtained from first principles,
as
$\alpha$.

How does this model produce a $\Lambda$ having mean zero and
fluctuations of the desired magnitude?
Owing to the random signs of the individual contributions to $S$,
their sum will vanish on average, but there will remain
residual fluctuations in $S$ of order $\alpha\sqrt{N}$.
Now according to one of the fundamental assumptions of causal set theory,
the volume of a spacetime region must be identified
--- modulo Poisson fluctuations ---
with the number of elements constituting that region,
i.e. $V\approx N$ in natural units.
Thus, the fluctuations in $S$ produce, at a given cosmic time,
a cosmological
term with a typical magnitude given, as desired, by
$\alpha\sqrt{N}/V \approx \alpha\sqrt{V}/V \approx \alpha/\sqrt{V}$,
a time-dependent value
which diminishes
as the volume of the
past grows.\footnote%
{The fluctuations are constrained by hand to be spatially homogeneous,
  this being probably the most serious limitation of the model as it has
  been developed so far.}
If we crudely identify $V$ with $(H^{-1})^4$,
$H$ being the Hubble parameter,
then we can see
with the help of the Friedmann equation
that one should expect
the magnitude of the fluctuations in
$\Lambda$
to track the total energy density:
$\Lambda \sim \sqrt{1/V} \sim 1/\sqrt{(H^{-1})^4} \sim H^2 \sim \rho_{critical}$.
And exactly this behaviour was observed in
the numerical simulations of \cite{everLam}.

Besides the restriction to spatial homogeneity,
the model of \cite{everLam} contains a second {\it ad hoc} element of
importance.
To understand where it comes from,
recall that
the cosmological term $\Lambda$ in the Einstein equation,
\begin{equation}\label{EE}
      {1\over\kappa} G^{ab} + \Lambda g^{ab} =  T^{ab}  \ ,
\end{equation}
has classically to be
a true constant unless $T^{ab}$ fails to be conserved
(or $\kappa$ fails to be a constant).\footnote
{Here $\kappa=8\pi G$, where $G$ is Newton's constant.
 Henceforth we take $\kappa=1$.}
In order to interpret
the predicted time dependence we must therefore depart from the
classical field equations.
To that end,
we first assume the universe to be homogeneous and isotropic,
so that (\ref{EE}) reduces
to a set of two ordinary differential equations
in a single dependent variable, the cosmological scale factor $a$.
For the case of a spatially flat universe,
which seems to be favored by the cosmological data,
these two equations are \cite{wein1}
\begin{equation}\label{moja}
  3H^2 - \Lambda = \rho                      
\end{equation}
and
\begin{equation}\label{mbili}
  {2\ddot{a} \over a} + {\dot{a}^2 \over a^2} - \Lambda = - p \ , 
\end{equation}
where a dot
denotes differentiation with respect to
proper time $\tau$,
$H = \dot a/a$ is the Hubble parameter,
$\rho$ is the energy density of non-gravitational matter
(including ``dark matter''),
and $p$ is its pressure.

For future reference,
notice that
if,
as is often done,
we interpret $-g^{ab}\Lambda$ as an effective addition to the
stress-energy $T^{ab}$ of perfect fluid form,
then (\ref{moja}) and (\ref{mbili})
entail
the identifications,
$\rho_\Lambda=\Lambda$ and
$p_\Lambda=-\Lambda$,
corresponding to the ``fluid equation of state'',
$\rho_\Lambda=-p_\Lambda$.

Of our two equations for $a$,
the first one
or {\it Friedmann equation}
is only of first order in time.  It is therefore a
constraint equation in the standard terminology,
and we will
refer to it in this paper as the {\it Hamiltonian constraint equation}
or HCEq.
The second equation is of second differential order,
and we will
refer to it as the {\it acceleration equation} or {\it AccEq}.

Now we wish to allow $\Lambda$ to vary with time.
Since doing
so
renders (\ref{moja}) and (\ref{mbili}) incompatible,
we can choose at most one of these equations as our
dynamical guide.
In reference \cite{everLam} we
let HCEq
play this role,
but that is not the only possibility,
albeit it is a natural choice and it
has the nice feature that one can
interpret the resulting scheme as a
{\it local}
change to
the ``equation of state of $\Lambda$''.\footnote
{That is, we can retain both Einstein equations
 if we treat the cosmological
 term as a perfect fluid, but make the altered
 identifications, $\rho_\Lambda=\Lambda$,
 $p_\Lambda = -\Lambda - \dot\Lambda/3H$.}
Nevertheless,
one would like to know how a
different choice would
affect our conclusions.  What if we based the model on equation (\ref{mbili})
rather than equation (\ref{moja}), or on some linear combination of the two,
such as the trace of the Einstein equation (TrEq),
which
also offers itself as
a natural choice  of dynamical guide?
To the extent that our phenomenological scheme
proved to be robust against such
modifications of its ad hoc elements,
we might feel more confident
that it adequately realized the underlying idea of $\Lambda$
fluctuations of typical magnitude $1/\sqrt{V}$.
Investigating this question of ``structural stability''
is
the purpose of
the present paper.

\section {The ``Mixed Equation'' and Structural Stability}

The nice results of \cite{everLam} were obtained from HCEq (\ref{moja}).
To what extent would they
persist if we used instead
an
arbitrary linear combination of HCEq with AccEq?

The first thing to notice in response to this question is that not all
linear combinations make sense, unfortunately, because AccEq (\ref{mbili}) is
inherently unstable, even with a non-fluctuating $\Lambda$
(indeed, even for $\Lambda=0$).
In the latter setting,
equation (\ref{mbili}) is actually redundant
since it is
(up to a factor)
just the time derivative of equation (\ref{moja}).
Conversely,
solving (\ref{mbili}) with initial data that
{\it exactly} satisfies (\ref{moja})
is equivalent to solving (\ref{moja}),
as is well known.
But if
one's
initial data fails
slightly to satisfy (\ref{moja}), the failure grows
with time.
This makes
(\ref{mbili})
unsuitable for numerical solution,
even when $\Lambda=0$.
Still less
would one expect it to be
suitable for a stochastic model like ours,
where random deviations
from the constraint submanifold are constantly being introduced
dynamically, even forgetting about
the
round-off and discretization errors of
one's
differencing scheme.\footnote
{The instability of the ``constraint submanifold''
 has often plagued attempts to solve the Einstein equations  numerically.}

The instability in question is brought out clearly if
one re-expresses \hbox{AccEq} in terms of the variable
$D = 3H^2 - \Lambda - \rho$,
which measures the degree to which
the HCEq fails to be satisfied.
Doing this yields for the time-dependence of $D$
\begin{equation}\label{nne}
  {\dot D \over 3H} + D = -{{\dot \Lambda} \over 3H}, 
\end{equation}
or
\begin{equation}\label{vier}
  a^{-3} {d \over d\tau} (a^3 D) = -{d\Lambda \over d\tau} \ , 
\end{equation}
%
while HCEq itself is, of course, equivalent to
\begin{equation}\label{tatu}
  D = 0 \ .                                     
\end{equation}
For a $\Lambda$ which doesn't fluctuate,
say for $\Lambda=0$,
the right hand side of equation (\ref{vier})
vanishes, and the time dependence of $D$ is particularly simple:
\begin{equation}\label{tano}
   a(\tau)^3 D(\tau) = constant, 
\end{equation}
or
\begin{equation}\label{sita}
  D(\tau) = \left({a(\tau_0) \over a(\tau)}\right)^3  D(\tau_0) \ .
\end{equation}
Noting that
by definition
$3H^2=\Lambda+\rho+D$,
and recalling that
an $a^{-3}$
scaling is precisely that of
``dust'' (pressureless matter),
we can interpret (\ref{sita}) as saying that
any deviation from  $D(\tau_0)=0$ acts as a
fictitious source of dust introduced at time
$\tau_0$.
In a radiation-dominated cosmos,
such a term
will ultimately
swamp
the genuine stress-energy,
no matter how small it starts,
since $\rho_{radiation}$ scales as $a^{-4}$.
At that point,
the relative error $D/\rho$ reaches $O(1)$ and
the behaviour of the solution becomes very
different from what it would have been had the error not been
introduced.  If $D_0$ happened to be negative, the
cosmos
would
re-contract, even though in the true solution it would have
continued expanding forever.
%
(Notice also that, for numerical purposes, the instability is much worse
than (\ref{sita}) would make it seem.  Because $a$ varies by a factor of
$10^{30}$ or so over the course of a simulation, it is impractical to
use $\tau$ as time-parameter.  Rather, something like $\log\tau$ must be
used, and with respect to such a time, the instability will be exponential.)

For a time-varying $\Lambda$ we find instead of (\ref{sita}),
\begin{equation}
  D(\tau) = \left({a(\tau_0)\over a(\tau)}\right)^3 D(\tau_0) +
         \int_{\tau_0}^{\tau} d\tau' \left({a(\tau')\over a(\tau)}\right)^3
         (-\dot\Lambda(\tau')),
\end{equation}
which says, in effect, that
a variation $d\Lambda$ introduces into the effective energy density a
fictitious dust contribution of $-d\Lambda$
(as if an amount $d\Lambda$ of $\Lambda$ had ``turned to dust'').
Although this expression is
less easy to analyze than (\ref{sita}),
it is hard to
believe that the fluctuations in $\Lambda$ under the integral sign could
avoid exciting the instability.
Indeed,
simulations we performed
with a fluctuating $\Lambda$
exhibited
an instability
in this case
that was as
bad as or worse than that for $\Lambda=0$.

As a check on this conclusion, we also simulated AccEq another way, which is
perhaps worth reporting here since it brings out the fact that our
scheme can be interpreted as a modification to the ``equation of state
of $\Lambda$'', rather than as a change to the Einstein equations.  In
this re-interpretation, we write the Einstein equations as
\[
  3H^2 = \rho_\Lambda  + \rho \ , \qquad
  {2\ddot{a}\over a} + H^2 = - (p_\Lambda + p) \ .
\]
The scheme based on (\ref{moja}) is then equivalent to the equation of state,
\begin{equation}\label{eosa}
  \rho_\Lambda = \Lambda \ ,
  \quad
  p_\Lambda =  -\Lambda - {d\Lambda/d\tau \over 3H} \ ,  
\end{equation}
according to which $p_\Lambda$ depends on both $\Lambda$ and its
time-derivative,\footnote
{An equation of state like this was considered in \cite{volovik}.
 Notice that, strictly speaking, $d\Lambda/d\tau$ is infinite because
 $\Lambda$ is (in the continuum limit) a function of Brownian type.}
%
while the scheme based on (\ref{mbili})
is equivalent to the equation of state
\begin{equation}\label{eosb}
  p_\Lambda = -\Lambda \ , \qquad
  \rho_\Lambda + \dot\rho_\Lambda / 3H = \Lambda, 
\end{equation}
according to which $\rho_\Lambda$ is not even a local (in $\tau$)
function of $\Lambda$.
As expected (since the two approaches are equivalent modulo numerical
errors) simulations with
this scheme
exhibited the same instability as
did direct simulation of equation (\ref{mbili}).

Now let us extend this stability analysis to
the equation formed as an arbitrary linear combination
(with coefficients $\nu$, $\mu$)
of HCEq (\ref{moja}) with AccEq (\ref{mbili}).
Expressing
the resulting equation
in terms of $D$ yields the corresponding linear
combination of (\ref{tatu}) and (\ref{nne}), namely
\begin{equation}\label{mix0}
  \mu {\dot D \over 3H} + (\mu+\nu)D = 0 \ , 
\end{equation}
where we have again taken $\Lambda=0$.
Instead of (\ref{tano}), we find now
\begin{equation}\label{mix00}
  D = {constant \over a^{3b}}  \ ,  
\end{equation}
where $b=(\mu+\nu)/\mu=1+\nu/\mu$.
In a radiation-filled cosmos, we therefore need $3b\ge4$,
in order that the error remain small.\footnote
{For completeness we mention that if later on
 the universe becomes ``matter dominated",
 since in that era mass density dilutes as $a^{-3}$,
 the inequality can be loosened slightly.}
The limiting case is thus $b=4/3$ or $\mu=3\nu$.
Interestingly, this corresponds exactly to
the trace equation, TrEq
(given by $\nu=1$, $\mu=3$, and equivalent to $G^a_a +4 \Lambda = T^a_a$),
which accordingly is ``marginally unstable''
in the structural sense.\footnote
{At the other end of the stable range, HCEq betrays no hint of
  instability when considered on its own.  But adding even a small amount
  of AccEq increases its differential order,  whence it is not strange
  that such a change can destabilize it when the sign of the addition is
  unfavorable.}
The trace equation (TrEq)
and the Friedmann equation (HCEq)
are then at the two extremes of the stable range,
and
one may expect that every other (stable)
combination will behave in a way intermediate between these two
extremes.
This makes TrEq all the more interesting,
supplementing its intrinsic interest in relation to conformal
variations of the metric.

In the rest of this paper,
we will refer to the above linear combination of AccEq with HCEq as
{\it MixedEq}, and we will limit $\mu/\nu$ to its stable range.
In terms of conformal time,
MixedEq is given by
 \begin{equation}\label{mix1}
 \mu \; a'' = {{\mu -3 \nu} \over 2} \, {(a')^2 \over a} +
            {a^3 \over 2} \, (\nu \, \rho_{total} - \mu \, p_{total}) +
            {a^3 \over 2} \, (\mu + \nu) \, \Lambda
\end{equation}
with
\[
   \rho_{total} = \rho_{radiation} + \rho_{matter} \ ,
\]
and
\[
  p_{total} = {\rho_{radiation} \over 3} \ .
\]
Here $\rho_{radiation/matter}$ is the energy density in
radiation/pressureless matter,
and
a prime represents
differentiation with respect to the conformal time,
$\eta=\int{d\tau/a}$.
For $\mu=3\nu$
we obtain the TrEq written as
\begin{equation}\label{trEq}
 {a'' \over a^3} = { {4\Lambda + \rho_{matter}} \over 6}\ . 
\end{equation}
(Notice that $\rho_{rad}$ drops out of TrEq, as it had to given the
conformally invariant nature of radiation.  For a radiation filled
universe with $\rho_{matter}=\Lambda=0$, (\ref{trEq}) yields simply
$da/d\eta$ = constant.)
We will see in the next section
that equation (\ref{mix1}) with $0<\mu\leq 3\nu$ is indeed stable
and produces results similar
to those obtained in reference \cite{everLam} from HCEq.

\section {Results of simulations}

Computer simulations of
MixedEq (\ref{mix1})
in the stable regime, $\mu \leq 3 \nu$,
produce results similar to those obtained
in \cite{everLam}
from
HCEq.
\begin{figure}[ht]
\includegraphics[scale= 0.95,origin=c]{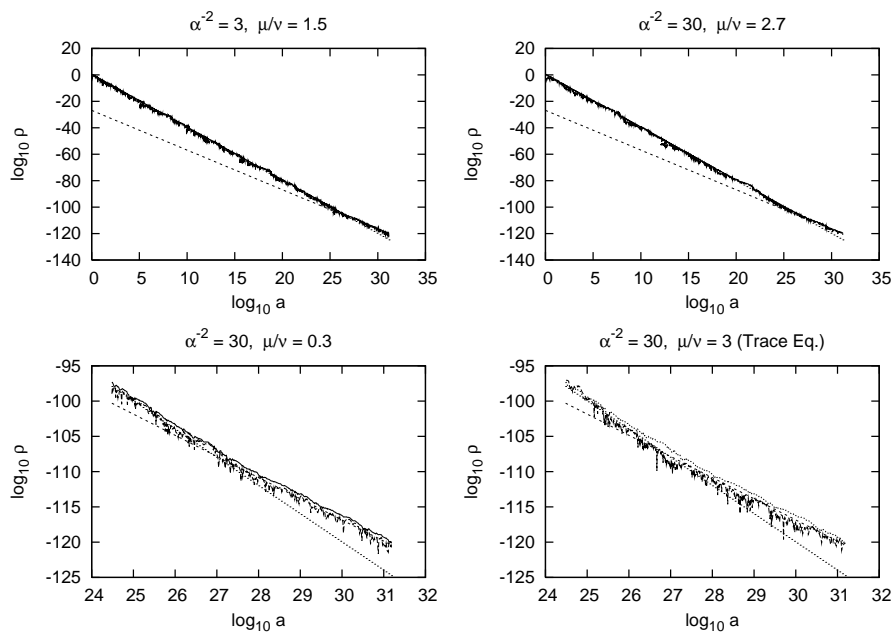}
\caption{Tracking behaviour exhibited by $|\Lambda|$  for
         various values of the model-parameters $\mu/\nu$ and $\alpha$. }
\label{f221}
\end{figure}
Figure \ref{f221}
displays
plots
of the energy density versus the scale factor in
some of these
 simulations,
and
the tracking behaviour alluded to earlier is clearly visible.
The top two diagrams
cover the whole period from the Planck time to the present epoch,
while
the bottom two diagrams focus on the transition
from $a^{-4}$ scaling (radiation domination, shown by the dotted line)
to $a^{-3}$ scaling (matter domination, shown by the dashed line)
of the ambient
energy density.
One sees that
the
effective
energy density in $\Lambda$ (solid jittery line)
switches its scaling as well and follows the
total effective energy density
(wiggly dotted line)
modulo fluctuations.
\begin{figure}[ht]
  \includegraphics[scale=0.9,origin=l, angle=-90]{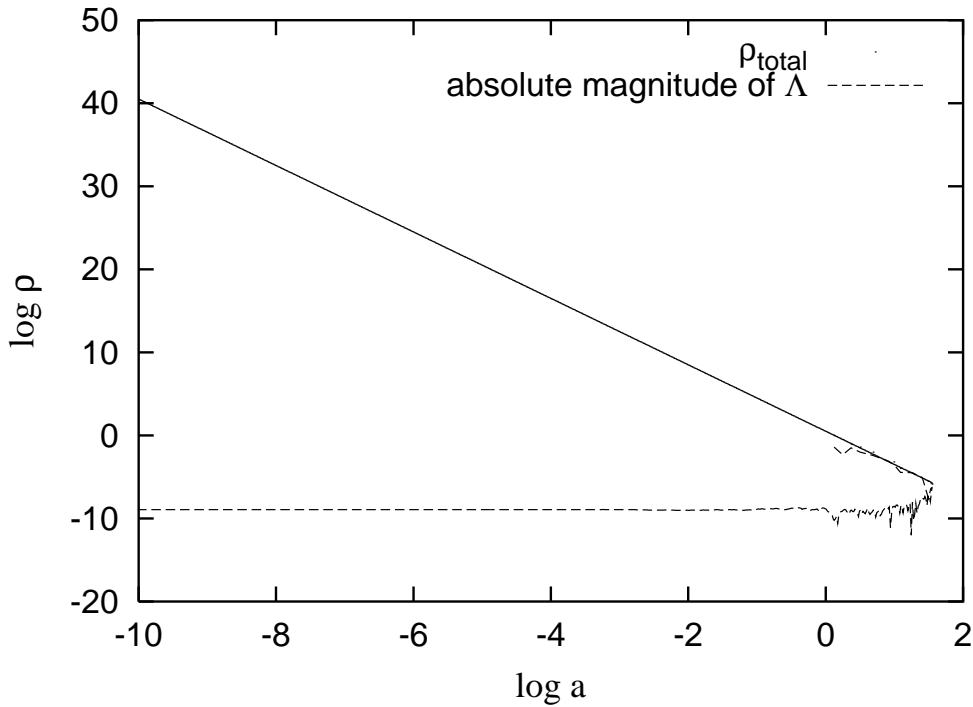}
  \caption
{Recollapse is typical of evolution with MixedEq (or TrEq), and prevents
  the cosmos from reaching its present size when $\alpha$ is too large.
  In the example shown $\alpha =1/5$.  The energy density in matter and
  radiation (solid line $\rho$) moves from left to right as the cosmos
  expands, then retraces its path, heading toward the final singularity.
  In contrast, $|\Lambda|$ (dashed line)
  tracks $\rho$ while the cosmos expands, but
  remains essentially frozen as it contracts.}
    \label{f23}
\end{figure}


One of the benefits of
using the MixedEq
(i.e., any of the stable combinations with non-vanishing $\mu$)
is the ability to
accommodate
larger values of the parameter $\alpha$.
With HCEq
almost all of
the simulations
terminate prematurely
when $\alpha$ is too big,
the reason being that
when $\alpha$ is large,
large fluctuations in $\Lambda$
are bound to happen, including negative ones.
If such a fluctuation overwhelms
$\rho$ in (\ref{moja}),
it renders
further evolution impossible by making $H$ imaginary.
There are intriguing suggestions of how to
keep going
in such a circumstance,
some of which were
discussed in \cite{everLam},
but a good understanding of their status is still lacking.\footnote
{The most obvious suggestion is simply to change the sign of $\dot{a}$
  at that point, but it seems that this alone cannot cure the problem in
  general.}

\begin{figure}[ht]
\includegraphics[scale=0.9,origin=c]{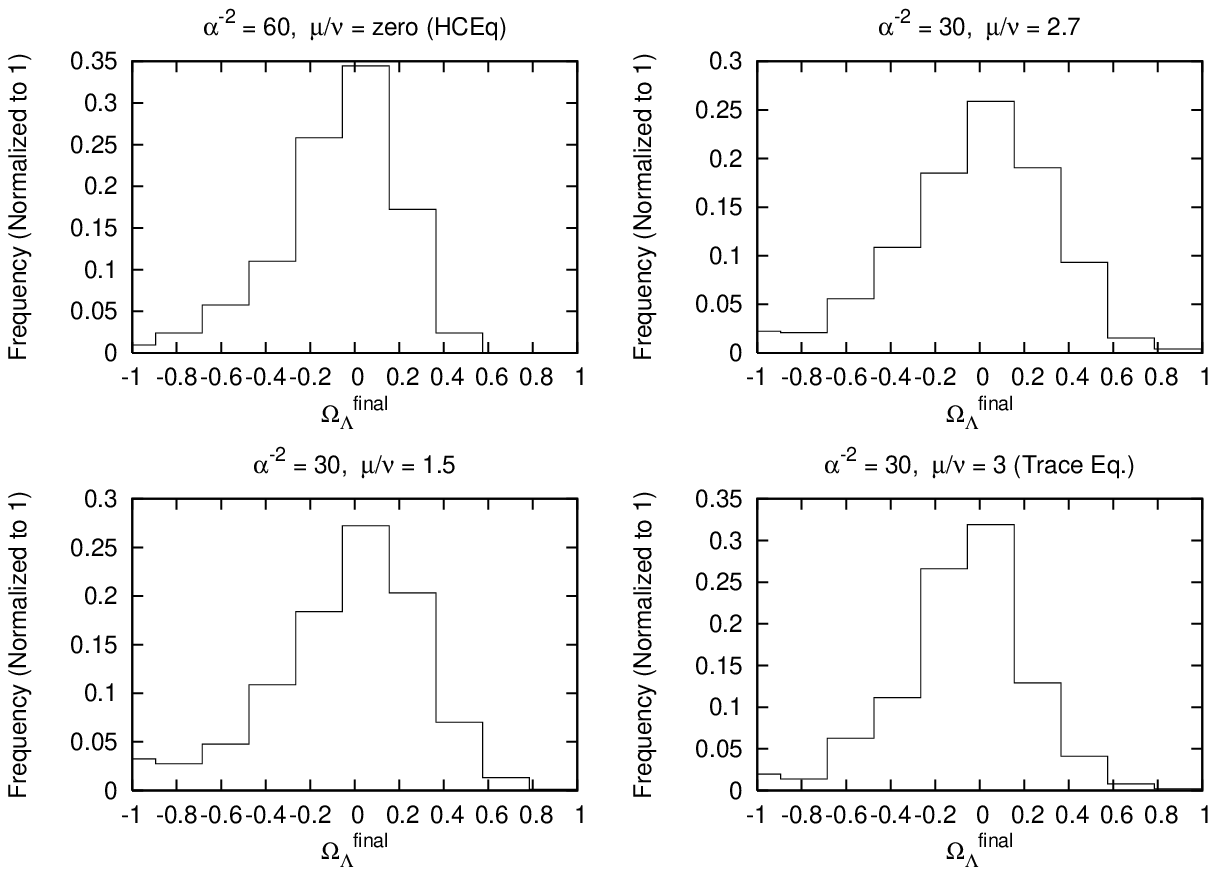}
\caption{Histograms of the final values of $\Omega_\Lambda$ for
         $\mu/\nu = 1.5,\, 2.7$ and $3$ ($\alpha = \sqrt{1/30}$).
       Also shown for comparison is a histogram for $\mu=0$ and $\alpha=\sqrt{1/60}$.}
\label{f222}
\end{figure}

On the other hand
for
MixedEq (\ref{mix1}) with $0 < \mu \leq 3 \nu$
a large negative fluctuation in $\Lambda$
simply means that $\ddot a$ becomes
negative.\footnote
{Actually $a''$ becomes negative but this implies that $\ddot a$ is also
 negative as can be seen from the equation
 $\ddot a = \frac{a''}{a^2}-\frac{a'^2}{a^3}$.}
This can eventually turn an expansion into a contraction
(and vice versa for a positive fluctuation),
but it
poses no problem of principle and it lets the simulation continue.
Thus we
have been
able to simulate
as high as $\alpha^2$ = $1/3$
with MixedEq,
whereas $\alpha^2 > 1/50$ was not
viable in \cite{everLam}.
The use of larger values of $\alpha$ has another, important benefit
as it
naturally results in larger final values of $\Omega_\Lambda$.
We will discuss this further
in connection with
figure \ref{f222}.

Although MixedEq permits us to evolve through a turning point,
thereby affording a larger latitude in the choice of $\alpha$,
simulations with $\alpha \gto 1$
still do not reach
the present value of the scale factor
(or equivalently the temperature)
in any significant number before re-collapsing.
Thus, the problem that affected HCEq
\cite{everLam}
shows up
in another guise with MixedEq,
albeit at larger values of $\alpha$.
The
biggest
$\alpha$
that
reached
the present
epoch
was $\sqrt{1/3}\approx 0.58$,
and
out of $10$ million runs
with this $\alpha$
only $25$ did so.
(The first graph in figure \ref{f221} shows one of these simulations).
In order to explain what goes wrong,
recall that the magnitude of $\Lambda$
in our model
decreases with the past volume (which, given our ansatz, is
constant on each hypersurface of homogeneity).
As long as the cosmos is expanding,
the absolute
magnitude of the $\Lambda$ term remains
comparable to
the energy density in radiation
(this being the relevant energy component at early times).
\footnote%
{This of course is the tracking behaviour that solves the ``Why Now?" puzzle.}
But when the cosmos begins to contract,
the
density of
radiation,
being proportional to $a^{-4}$,
increases sharply.
And
since the past-volume keeps
on accumulating,
after a while the $\Lambda$ fluctuations are too small to
reverse the
sign of $\dot a$,
and
the probability
for the universe to recover from the collapse
practically vanishes.
The graph of Figure \ref{f23} illustrates one such case.

\begin{figure}[ht]
\includegraphics[scale=0.8,origin=c]{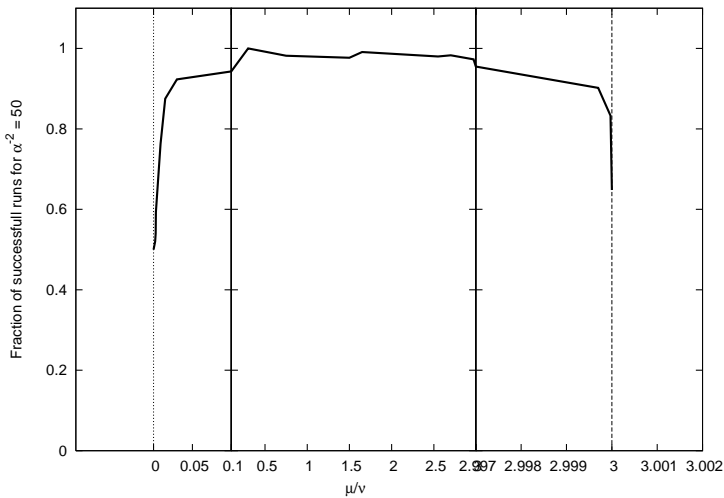}
\caption
{The fraction of runs that reach the present time,
 plotted as a function of $\mu/\nu$, for $\alpha^{-2} = 50$.
 The HCEq corresponds to $\mu/\nu=0$, the TrEq to $\mu/\nu=3$.}
\label{f226}
\end{figure}
\begin{figure}[ht]
\includegraphics[scale=0.8,origin=c]{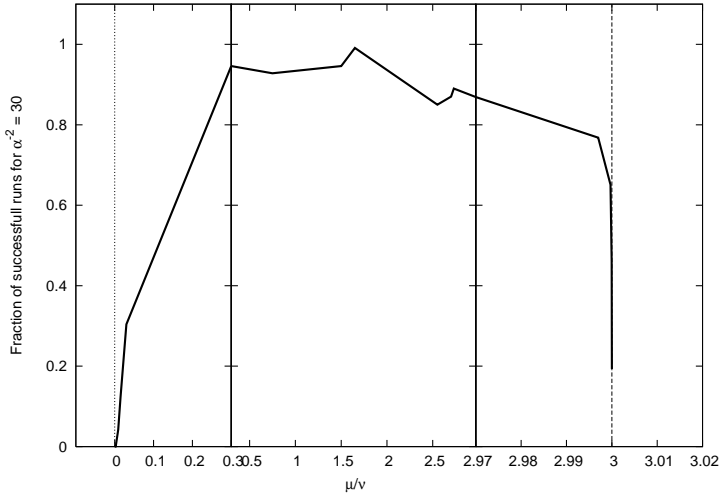}
\caption
{Same as figure \ref{f226} but for $\alpha^{-2} = 30$.}
\label{f227}
\end{figure}
\begin{figure}[ht]
\includegraphics[scale=0.8,origin=c]{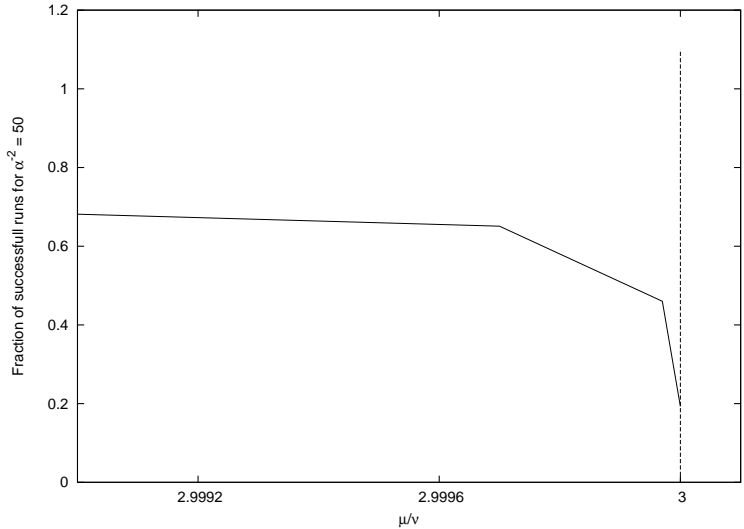}
\caption
{The end part of the curve in figure \ref{f226} magnified about twenty times.}
\label{f28}
\end{figure}

As an indicator of
how our model reacts
as we move away
from the HCEq ($\mu = 0$)
into the MixedEq $(0 < \mu < 3\nu)$
and then reach the other extreme at the TrEq $(\mu = 3\nu)$,
we studied the fraction of times
our simulations reached the present value of the
scale-factor.
For $\alpha^2 = 1/30, 1/50$, and $1/70$,
we simulated HCEq, TrEq,
and MixedEq
with values of $\mu / \nu$
as small as $0.002$ (just after the HCEq)
and reaching as close as $2.99998$ to the TrEq ($\mu/\nu = 3$).
Figures \ref{f226} and \ref{f227}
plot this
fraction as
a function of $\mu / \nu$ for $\alpha^2 = 1/50$,
and $1/30$,
respectively.
In both plots, one observes a central ``plateau'' in which most
realizations reach today's
scale-factor,
flanked on each side by
``escarpments'' in which the fraction of such realizations falls
precipitously as one approaches the unstable range.
Close to the TrEq the
dependence
seems to be exponential,
so
in figure \ref{f28}
we have expanded that part of the curve for $\alpha^2 = 1/50$
to reveal the continuous transition to $\mu/\nu = 3$.

Figure \ref{f222} shows
some
histograms
of the final values of $\Omega_\Lambda$
(called $\Omega_\Lambda^{final}$)
for HCEq with $\alpha^2 = 1/60$
and
for MixedEq with $\alpha^2 = 1/30$,
and
for
various
$\mu / \nu$ values.
It can be seen that
the values of $\Omega_\Lambda^{final}$
are nicely peaked at zero
and fall off on either side as expected.
The first
thing to notice is that
$\Omega_\Lambda^{final}$ goes
as high as $0.99$.
About $5$ percent of
the time,
it lies
above $0.5$,
and about $2$ percent above $0.7$.
The positive values
of $\Omega_\Lambda^{final}$
slightly outweigh the negative
ones,
the probable reason being
that a universe which spends
more of its time with positive $\Lambda$ is
more likely to reach the present epoch without recollapsing.
In general the positive values seem to make up
fifty to sixty
percent of the total.
One more thing that
catches the eye is
that
the tail on the negative side
is longer than
on the positive side.
This
merely reflects
the
asymmetry
built into
the definition,
$\Omega_\Lambda \equiv \rho_\Lambda / (\rho_\Lambda + \rho_m)$,
in consequence of which
$\Omega_\Lambda$
normally\footnote
{It can happen that $\rho_\Lambda + \rho_m$ becomes
 negative such that $\Omega_\Lambda > 1$.  However the re-contraction
 that accompanies a
 negative
 net effective energy density enhances the matter-density,
 and $\Omega_\Lambda$ rapidly returns to its ``normal'' range, either
 for this reason or because
 $\Lambda$ fluctuates back to a positive value.}
%
is bounded by
$+1$ on the positive side,
while ranging
to $-\infty$ on the negative side.
(A few percent of the $\Omega_\Lambda$ fell below $-1$, but we did not
plot them on the histograms.)

\section {Conclusions}

This study is a follow-up to reference \cite{everLam},
where we
recalled how
the causal set idea
leads to fluctuations in the cosmological term
and showed that if
the mean value
of $\Lambda$ is taken to be zero then these fluctuations
have the potential to account for the presently observed $\Omega_{\Lambda}$.
We also raised some concerns about the model, namely:

\begin{itemize}

\item The nice results of \cite{everLam} were obtained from the choice
  of equation (\ref{moja}) as ``dynamical guide'', and it was
  uncertain
  whether (\ref{mbili}) or a linear combination of the two equations
  would behave similarly.

\item It was not clear how to continue the evolution when the total
  effective energy density became negative.

\item The parameter $\alpha $ that governed the magnitude of the
  fluctuations had to be five to ten times smaller than
  unity, and this looked like a mild fine tuning problem.

\item Rarely did the model produce a final value for $\Omega_{\Lambda}$
large enough to accord with present observations.
\end{itemize}

\noindent
The present work throws light on all of these
issues.
We take them in turn.

\begin{itemize}

\item
It turns out that the second order equation (\ref{mbili}),
referred to in this paper as the AccEq,
is inherently unstable and cannot replace HCEq (\ref{moja}) in the model
of \cite{everLam}.
However a large range of convex combinations of the two equations
is stable, specifically combinations of the form
$\nu\times{\rm HCEq}+\mu\times{\rm AccEq}$ with $0 \leq \mu \leq 3\nu$.
In our simulations, all of
these stable combinations
exhibited the same tracking behaviour as was
obtained in \cite{everLam} from the purely first-order HCEq ($\mu =0$).
Our question about ``structural stability''
is therefore answered in the affirmative.
\item When the net effective energy-density of a universe
(evolving under one of the stable mixtures of HCEq and AccEq)
becomes sufficiently negative,
an
expanding universe
starts to contract.
In principle, it can re-expand if the net effective energy-density
becomes positive again,
but
this
happens only rarely.
In any case, there is for $\mu>0$,
no problem in following an expansion and
re-contraction all the way down to a singularity of infinite density.
\item
Whereas
in \cite{everLam}
with the HCEq,
we were only able to
simulate
as high as $\alpha^2 \approx 1/50$,
we
have been able
with MixedEq to
go as high as $\alpha^2 = 1/3$,
albeit
$1/30$ is a more realistic upper limit
if the universe is to
reach its present size
with an appreciable probability.
\item
Since
with MixedEq
we can handle larger values of $\alpha$
(with their concomitant larger fluctuations),
we can end up with larger
final values of $\Omega_{\Lambda}\,$.
For $\alpha \sim 1/30$
we can account for
the present observational value
of $\Omega_{\Lambda}$
in about 2 percent of the
simulations.

\end{itemize}

\noindent
It is important to
realize,
however,
that our
variable
$\Omega_{\Lambda}$
cannot be
compared directly
with observational parameters quoted in the literature which assume that
$\Lambda$ is constant (or has an ``equation of state'' with
constant $w=p_\Lambda/\rho_\Lambda\,$).
When instead, $\Lambda$ is fluctuating,
the comparison with observation
must be derived anew.
To that end, one might for example
use our simulations to construct fictitious
data sets
of
supernova luminosity and redshift,
one for each run of the simulations.
One would then feed these data sets
into one of the algorithms people have used to obtain the quoted values of
$\Lambda$ and $w$, and one would ask how often the resulting values came
near to the quoted ones.

Can the accelerating Hubble expansion be traced to a fluctuating
$\Lambda$ and can such fluctuations be understood as a nonlocal and
quantal residue of an underlying spatio-temporal discreteness?

This long-standing idea has still to be embodied in a fully fledged
phenomenological model, but an encouraging start was made in
\cite{everLam}.
The model of \cite{everLam} involved some ad hoc choices, however,
together with an artificial restriction to spatial homogeneity.
In this paper we have not dealt with the latter shortcoming\footnote%
{The consequent risk to the model was pointed out in \cite{everLam} and
  emphasized further in \cite{barrow}.}
but have
concentrated on the ambiguity inherent in the manner in which
\cite{everLam} implemented the idea of a varying $\Lambda$.  So far, the
evidence is that none of the qualitative features of the model depend on
how that ambiguity is resolved, and overall our results sustain the
picture developed in \cite{everLam}, according to which the fluctuating
$\Lambda$ tends to remain, throughout the phase of cosmic expansion, in
rough equilibrium with the ambient matter density.

There remains,
though, a kind of tension between the magnitude of the fluctuations (as
reflected in the parameter $\alpha$) and the continuation of the
expansion.  If $\alpha$ is too big, the negative fluctuations will tend
to terminate the expansion; if it is too small, the positive
fluctuations will be unable to account for the current value of
$\Lambda$.
If one chooses $\alpha$ appropriately, the two competing effects can be
balanced\footnote%
{Order of magnitude estimates suggest that this balancing is possible
  only in $3+1$ dimensions, as pointed out also in \cite{barrow}.
}
fairly well, but some discrepancy always seems to remain.
Indeed, it is likely a generic feature of the models based on MixedEq
that for any choice of $\alpha$,
the cosmos
given long enough,
will eventually recollapse due to a
negative fluctuation.

If this is so, we hope we may be excused for speculating further that cycles of
expansion and contraction would succeed each other indefinitely, their
characteristic lifetime depending on the value of $\alpha$ as it emerges
at the start of each new cycle.  In this way, the ``Tolman-Boltzmann''
scenario
of \cite{peyresq}  would have acquired a possible
 basis in quantum gravity.
Although that scenario emerged from a non-quantum dynamics for causal
sets, namely that of
{\it classical sequential growth},
it illustrated how
the most striking features of our universe --- its approximate spatial
homogeneity and isotropy --- might have emerged dynamically over the
course of repeated expansions and collapses.
There the collapses occurred via
rare
statistical fluctuations following exponentially long
periods of stasis.  Here, the collapses would
be more dynamical in character, initiated by
{\it quantal} fluctuations in $\Lambda$, or
in other words fluctuations in
the form of the gravitational field equations.

This research was supported in part by NSERC through grant RGPIN-418709-2012.
Research at Perimeter Institute for Theoretical Physics is supported in
part by the Government of Canada through NSERC and by the Province of
Ontario through MRI.

\end{document}